\documentstyle[prd,epsf,aps,floats]{revtex}

 \newcommand{\Sp}{\mbox{Sp}}

\draft

\begin{document}
\twocolumn[\hsize\textwidth\columnwidth\hsize\csname
@twocolumnfalse\endcsname
\title{%
\hbox to\hsize{\normalsize\rm E-print hep-ph/9903415
\hfil Preprint YARU-HE-99/02}
\vskip 40pt
Axion in an External Electromagnetic Field}

\author{N.V.~Mikheev, A.Ya.~Parkhomenko, and L.A.~Vassilevskaya}
\address{Yaroslavl State (Demidov) University, Sovietskaya 14,
         Yaroslavl 150000, Russia}

\date{\today}

\maketitle

\begin{abstract}
We investigate the effective interaction of a pseudoscalar
particle with two photons in an external electromagnetic field
in the general case of all external particles being off the mass shell.
This interaction is used to study the radiative decay $a \to \gamma \gamma$
of the axion, a pseudoscalar particle associated with the
Peccei-Quinn symmetry, and the crossing channel of photon splitting
$\gamma \to \gamma a$. It is shown that the external field removes 
the suppression  associated with the smallness of the axion mass 
and so leads to the strong catalysis of these processes. 
The field-induced axion emission by the photon is analyzed as one more 
possible source of energy losses by astrophysical objects.
\end{abstract}
\vskip2.0pc]


\section{Introduction}

For quite a long time there has been interest in various extensions
of the standard model (SM) with a nonstandard set of Higgs particles.
In the simplest version of the SM one needs only one Higgs doublet.
However, there are reasons to believe that if scalar bosons
exist at all, their number may be significant.
The Higgs sector can be so rich that a situation becomes
possible when one or several Goldstone degrees of freedom
are not absorbed by the Higgs mechanism and appear as
physical massless particles. Such Goldstone (or pseudo Goldstone)
bosons, whose existence is related to the breakdown of exact (approximate)
symmetries, is also of interest from the experimental point of view:
they could be observed in low-energy experiments. Such particles include,
e.g., axions~\cite{Peccei77,WW}, Majorons~\cite{Maj}, arions~\cite{AU}.
The most widely discussed pseudo Goldstone boson, the axion, associated
with the Peccei-Quinn (PQ) symmetry continues to be an attractive
solution to the problem of $CP$ conservation in strong
interactions~\cite{Peccei96}.
Although the original axion, associated with the spontaneous breakdown
of the global PQ symmetry at the weak scale, $f_w$,
is excluded experimentally, modified versions of
PQ models with their associated axions are of great interest.
If the breaking scale of the PQ symmetry, $f_a$, is much larger than
the electroweak scale $f_a \gg f_w$, the resulting ``invisible
axion'' is very weakly interacting (coupling constant $\sim f_a^{-1}$),
very light ($m_a \sim f_a^{-1}$) and very long lived. The axion
lifetime in vacuum is gigantic:
\begin{equation}
\tau (a \to 2\gamma) \sim 6.3 
\times 10^{48} \, \mbox{s} \,
\left ( \frac{10^{-3} \, \mbox{eV}}{m_a} \right )^6 \;
\left ( \frac{E_a}{1 \, \mbox{MeV}} \right ) .
\label{eq:T0}
\end{equation}

The allowed range for the axion mass
is strongly constrained by astrophysical and cosmological
considerations which leave a rather narrow window
~\cite{Turner,Raffelt90,Raffelt-castle97,Raffelt-school97}\footnote{
However in Ref.~\cite{Rubakov}
a possibility to solve the $CP$ problem of QCD within a grand unified 
theory model with a heavy axion $m_a \alt 1$~TeV is considered.}: 
\begin{equation} 
10^{-5} \, \mbox{eV} \alt m_a \alt 10^{-2} \, \mbox{eV}, 
\label{eq:MsAx}
\end{equation}
where axions could exist and provide a significant fraction or
all of the cosmic dark matter. A survey of various processes involving
the production of weakly coupled particles and astrophysical methods for
obtaining constraints on the parameters of axion models is given
in Ref.~\cite{Raffelt-book}.

The important role which axions play in elementary particle physics
stimulates constant interest in theoretical and experimental studies.
Along with laboratory searches~\cite{Kyoto,Livermore}, stars are a unique
laboratory to investigate physical processes~\cite{Raffelt-book}.
For example, the upper bound on $m_a$ is obtained from the requirement
that stars not lose too much energy by axions.
In studies of processes occurring inside astrophysical
objects one has to take into account that dense matter substantially
influences particle properties.
An intensive electromagnetic field also plays the role of an
active medium modifying the dispersion relations of particles
so that novel processes are not only opened kinematically,
but become substantial as well. Photon splitting
$\gamma \to \gamma \gamma$~\cite{Adler71} and the Cherenkov process
$\nu \to \nu \gamma$~\cite{Raffelt97,Ljuba97} are among them.

Actually, both components of the active medium, plasma and
a magnetic field, are presented in most astrophysical objects.
A situation is also possible when the magnetic component dominates.
For example, in a supernova explosion or in a coalescence of neutron
stars a region outside the neutrinosphere of order of several tens
of kilometers with a strong magnetic field and a rather rarefied plasma
could exist. The possible existence of astrophysical objects with
$B \sim 10^{15} - 10^{17}$~G, significantly above the critical, Schwinger
value $B_e = m^2_e / e \simeq 4.41 \times 10^{13}$~G, was discussed both
for toroidal~\cite{toroidal} and for poloidal~\cite{poloidal} fields.

In this paper we investigate a field-induced effective interaction 
of a pseudoscalar particle with two photons described by three-point 
loop diagrams. By now the detailed studies of the analogous 
three-point loop vertex of the photons' interaction
in external electromagnetic fields are known only.
The history of these investigations
goes back to the pioneer paper by Adler~\cite{Adler71}
and it is still in progress~\cite{Ritus,Baier,Adler96,ChKM}.

Note that amplitudes of processes in an external field depend
not only on the kinematical invariants of type $m^2$ or $p^2$,
but also on field invariants $\vert e^2 (F F) \vert^{1/2}$,
$\vert e^2 (\tilde F F) \vert^{1/2}$, $\vert e^2 (p F F p) \vert^{1/3}$
where $m$ and $p_\mu$ are the mass and the four-momentum of a
particle, $F_{\mu \nu}$ is the external field tensor,
and $ (p F F p) = p^\mu F_{\mu \nu} F^{\nu \rho} p_\rho$.
In the ultrarelativistic limit the field invariant
$\vert e^2 (p F F p) \vert^{1/3}$ can occur as the largest one.
This is due to the fact that in the relativistic
particle rest frame the field may turn out to be of order of the
critical one or even higher, appearing very close to the constant
crossed field (${\bf E} \perp {\bf B}$, $E = B$), where
$(FF)=(F \tilde F) = 0$. Thus, calculations in this field
are the relativistic limit of calculations in an arbitrary
relatively smooth field.

In Sec.~II we present the result of our calculations of the 
$a \gamma \gamma$ vertex in the external crossed
field in a general case when the external particles are off
the mass shell. 
In Sec.~III we discuss photon and axion dispersion relations
in the crossed field and analyze the influence of the field-induced
``effective masses'' of the particles on kinematics.
It is shown that a novel channel of the photon decay 
$\gamma \to \gamma + a$ is opened kinematically.
In Sec.~IV the effective $a \gamma \gamma$ vertex is used to study
the axion decay into two photons $a \to \gamma + \gamma$.
A catalyzing influence of the external field on this
process is analyzed.
In Sec.~V we study the photon decay $\gamma \to \gamma + a$.
The energy loss by a photon gas due to this process is investigated
as an additonal source of the axion luminosity in a supernova explosion.
In the appendixes, the important details of the calculation of 
the $a \gamma \gamma$ vertex are collected.

\section{$\lowercase{a} \gamma \gamma$ vertex in the crossed field}

Invisible axion models are classified into two types,
depending on whether or not axions couple to leptons
(see, e.g.,~\cite{Peccei96,Raffelt-book}).
Here we investigate the field-induced $a \gamma \gamma$ interaction
of ``Dine-Fischler-Srednicki-Zhitnitsky (DFSZ) axions''~\cite{DFSZ} 
which couple with both quarks and leptons at the tree level. 

The Lagrangian describing the axion-fermion interaction is 
\begin{equation}
{\cal L}_{af} =  - i g_{af} (\bar f \gamma_5 f) a,
\label{eq:L1}
\end{equation}
where $g_{af} = C_f m_f/f_a$ is a dimensionless
Yukawa coupling constant, $C_f$ is a model-dependent
factor~\cite{Raffelt-book}, $m_f$ is the fermion's mass,
$\gamma_5$ is the Dirac $\gamma$ matrix, 
and $a$ and $f$ are the axion and fermion fields. 

\begin{figure}[tb]
\centerline{\epsfxsize=.4\textwidth \epsffile[215 410 455 520]{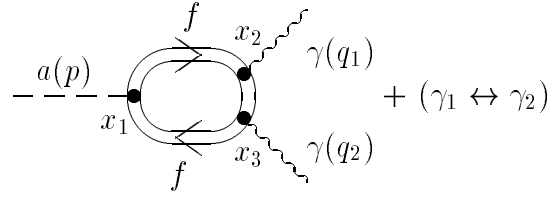}}
\caption{The effective vertex of the interaction
         of a pseudoscalar particle with two photons.
         Double lines correspond to fermion propagators
         in an external electromagnetic field.}
\label{fig:diag1}
\end{figure}

The two-photon axion interaction is investigated
in a general case when the photons are not assumed to be on the mass
shell. It means that the expression for the $a \gamma\gamma$ vertex
can be used as an effective Lagrangian of axion-photon interaction
in studies of processes involving axion or other pseudoscalar
particles with a coupling of type~(\ref{eq:L1}).
In the third order of the perturbation theory the matrix element 
of the effective $a \gamma \gamma$ interaction in an external field 
is described by two three-point loop diagrams (Fig.~\ref{fig:diag1}), 
where double lines imply that the influence of the external field
in the propagators of virtual charged fermions 
is taken into account exactly.

%
%
%

The general form for the matrix element corresponding to the diagrams
in Fig.~\ref{fig:diag1} is 
\begin{eqnarray}
S & = & {  4 \pi \alpha Q_f^2 g_{af} \over
\sqrt {2 E_a V \times 2 \omega_1 V \times 2 \omega_2 V}}
\nonumber \\
& \times &
\int d^4 x_1 \, d^4 X \, d^4 Y \,
\exp \left [ i (x_1 (q_1 + q_2 - p)) \right ]
\nonumber \\
& \times &
\exp \left ( - i
\left \lbrace
(q_1 X) + (q_2 Y) + \frac{e Q_f}{2} (X F Y)
\right \rbrace
\right )
\label{eq:S1} \\
& \times & \Sp \lbrace S(Y) (\varepsilon_2 \gamma )
S(X - Y) (\varepsilon_1 \gamma ) S(-X) \gamma_5 \rbrace
\nonumber \\
& + &  ( \varepsilon_1, q_1 \leftrightarrow \varepsilon_2, q_2 ),
\nonumber
\end{eqnarray}
where $\alpha = 1/137$ is the fine-structure constant;
$e > 0$ is the elementary charge,
$Q_f$ is a relative fermion charge in the loop;
$p$ is the four-momentum of the decaying axion,
$q_{1,2} = (\omega, {\bf q})_{1,2}$ and $\varepsilon_{1,2}$ are
the four-momenta of the final photons and their polarization four-vectors,
respectively; $(\varepsilon \gamma ) = \varepsilon_\mu \gamma^\mu$,
$\gamma_\mu$ are the Dirac $\gamma$ matrices; $X = x_1 - x_2$,
$Y = x_1 - x_3$; $S (X)$ is the translationally invariant
part of the fermion propagator in the crossed field
(Appendix~\ref{app:propagator}).

Further we will calculate a vertex function $\Lambda$ connected
with the matrix element Eq.~(\ref{eq:S1}) in the following way:
\begin{eqnarray}
S = {i (2 \pi)^4 \,\delta^{(4)} ( p - q_1 - q_2 ) \over
\sqrt {2 E_a V \times 2 \omega_1 V \times 2 \omega_2 V}} \, \Lambda.
\label{eq:S0}
\end{eqnarray}
The function $\Lambda$, thus defined, is the effective Lagrangian
of the axion-photon interaction in the momentum representation.
The calculation technique of the three-point vertex $a \gamma \gamma$
in the crossed field is given in Appendixes~\ref{app:bable}
and~\ref{app:vertex}.
In the following, we will consider only the field-induced contribution
$\Lambda^{(F)} = \Lambda - \Lambda^{(0)}$, where $\Lambda^{(0)}$
is the vacuum part of $\Lambda$,
to the effective Lagrangian, corresponding to the
transition $a(p) \to \gamma(q_1) + \gamma(q_2)$:
\begin{eqnarray}
\Lambda^{(F)} & = &
\frac{\alpha}{\pi} \sum_f \frac{Q^2_f g_{af}}{m_f}
  \Big [ (f_1 \tilde f_2) J_0 +
 (f_1 \tilde{\cal F} f_2) J_{1}
 \nonumber \\
& + &  (f_1 \tilde{\cal F})(f_2 {\cal F}) J_2
+ (f_2 \tilde{\cal F})(f_1 {\cal F}) \tilde J_2
\label{eq:V1} \\
& + & (f_1 \tilde{\cal F}) \frac{(q_2 f_2 {\cal F} {\cal F}q_2)}
{(q_2 {\cal F} {\cal F} q_2)} J_{3}
+ (f_2 \tilde{\cal F}) \frac{(q_1 f_1 {\cal F} {\cal F} q_1)}
{(q_1 {\cal F}{\cal F} q_1)} \tilde J_{3}
  \Big ] ,
 \nonumber
\end{eqnarray}

\begin{eqnarray}
{\cal F}_{\mu\nu} & = & \frac{F_{\mu\nu}}{F_f}, \,\,\,\,
{\tilde{\cal F}_{\mu \nu}} = \frac{1}{2} \,
\varepsilon_{\mu \nu \alpha \beta}
{\cal F}_{\alpha \beta},  \,\,\,\,
F_f  =  \frac{m^2_f}{e Q_f},
\label{eq:Ten} \\
f_{i \alpha \beta} & = & q_{i\alpha} \varepsilon_{i\beta} -
q_{i\beta} \varepsilon_{i\alpha} ,
\,\,
\tilde f_{i \alpha \beta} = \frac{1}{2} \,
\varepsilon_{\alpha \beta \mu \nu} \, f_{i \mu \nu} ,
\,\, i = 1, 2,
\nonumber
\end{eqnarray}

\begin{eqnarray}
J_0 & = & i \int\limits_0^\infty dz  \int\limits_D dx dy
 \left [ \cos \varphi_2 \, e^{- i \Phi} -
 e^{- i z \varphi_0 } \right ],
\nonumber \\
J_{1} & = & 2 \int\limits_0^\infty z dz \int\limits_D dx dy \,
\nonumber \\
& \times &
 \left [ 1 -
 \left ( 1 + \frac{\chi}{\chi_1} \right ) \, x -
 \left ( 1 + \frac{\chi}{\chi_2} \right ) \, y \right ] \,
\sin \varphi_2 \, e^{- i \Phi} ,
\nonumber \\
J_{2} & = & \frac{i \chi}{\chi_2} \,
\int\limits_0^\infty z^2 dz \int\limits_D dx dy \,
 y ( 1 - 2 x - y ) \, \cos \varphi_2 \, e^{- i \Phi},
\nonumber \\
J_{3} & = & \int\limits_0^\infty z dz \int\limits_D dx dy \,
 ( 1 - 2x ) \, \sin \varphi_2 \, e^{- i \Phi},
\label{eq:V2} \\
\tilde J_{i} & = & J_{i} (q_1 \rightarrow q_2), \qquad i = 2, 3,
\nonumber \\
\Phi & = &  z \varphi_0 + \frac{z^3}{3} \varphi_1,
\nonumber \\
\varphi_0 & = & 1 - 2xy \, \frac{(q_1 q_2)}{m^2_f}
- y (1 - y) \, \frac{q^2_1}{m^2_f} - x (1 - x) \, \frac{q^2_2}{m^2_f},
\nonumber \\
\varphi_1 & = & \left [ x (1 - x) \vert \chi_2 \vert -
y (1 - y) \vert \chi_1 \vert \right ]^2
\nonumber \\
& + & 2 x y (1 - x - y)^2 ( \vert \chi_1 \chi_2 \vert - \chi_1 \chi_2 )
\nonumber \\
& + & 2 x y [ xy + (x + y) (1 - x - y)]
(\vert \chi_1 \chi_2 \vert + \chi_1 \chi_2 ) \ge 0,
\nonumber \\
\varphi_2 & = & - 2 z^2 xy (1 - x - y) \,
\frac{(q_1 {\cal F} q_2)}{m^2_f},
\nonumber \\
\chi^2 & = & \frac{(p {\cal F} {\cal F} p)}{m_f^2}, \qquad
\chi_i^2 = \frac{(q_i {\cal F} {\cal F} q_i)}{m_f^2},\;\;\; i = 1,2.
\nonumber
\end{eqnarray}
The integration area $D$ in the integrals $J_i$ is $x, y \ge 0$,
$x + y \le 1$.

\begin{figure}[tb]
\centerline{\epsfxsize=.4\textwidth \epsffile[200 385 440 490]{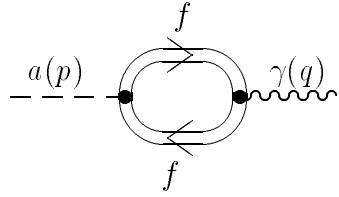}}
\caption{The diagram describing the effective axion-photon
         interaction.}
\label{fig:axgam}
\end{figure}

We note that the vertex $\Lambda^{(F)}$~(\ref{eq:V1}) contains
as a limit an amplitude of the field-induced effective
$a \gamma$ interaction which can be used in studies of axion-photon
oscillations. The amplitude of the $a(p) \to \gamma(q)$ transition
(the diagram is shown in Fig.~\ref{fig:axgam})
%
%
%
%
is obtained from Eq.~(\ref{eq:V1}) by means of replacing the field
tensor of one of the photons by the electromagnetic field tensor:
\begin{eqnarray}
 q_{1 \alpha} \to  q_{\alpha}, \;
 f_{1 \alpha \beta} \to f_{\alpha \beta}, \;
 q_{2 \alpha} \to 0, \;
 f_{2 \alpha \beta} \to - i F_{\alpha \beta}.
\nonumber
\end{eqnarray}
As a result, the amplitude of the $a \to \gamma$ transition
can be presented in the form~\footnote{For a numerical evaluation
note that the external field tensor $F_{\mu \nu}$ is presented
here in rationalized natural units.}:
\begin{equation}
M = - {i \alpha \over 2 \pi} \, ( f \tilde F )
\sum \limits_f \,{Q_f^2 \, g_{af} \over m_f } \,
\left [ I - I (F = 0) \right ] ,
\label{eq:MT}
\end{equation}

\begin{eqnarray}
I & = & \left ( {4 \over \chi} \right )^{2/3} \,
\int \limits_0^{\pi/2} \;
f(\eta) \, \sin^{- 1/3} \phi \, d\phi,
\nonumber \\
f(\eta) & = & i \,\int \limits_0^\infty du \, \exp
\left \lbrace
-i \, \left ( \eta u + { u^3 \over 3 } \right )
\right \rbrace ,
\nonumber \\
\eta & = & \left ( 1 - \frac{q^2 \sin^2 \phi}{4 m_f^2} \right )  \;
 \left ( {4 \over \chi \sin^2 \phi} \right )^{2/3} ,
\nonumber
\end{eqnarray}
where $f (\eta)$ is the Hardy-Stokes function and
$\chi$ is the dynamic parameter of the axion
defined in Eq.~(\ref{eq:V2}).

\section{Field-induced ``effective masses'' of the particles}

The vertex $a \gamma \gamma$ Eq.~(\ref{eq:V1})
can be used as an effective axion-photon
Lagrangian in studies of processes such as the radiative
decay $a \to \gamma \gamma$, photon splitting $\gamma \to \gamma a$,
and the Primakoff-type process for photoproduction of pseudoscalars
on electrons $\gamma + e \to e + a$.

\begin{figure}[tb]
\centerline{\epsfxsize=.4\textwidth \epsffile[225 320 425 610]{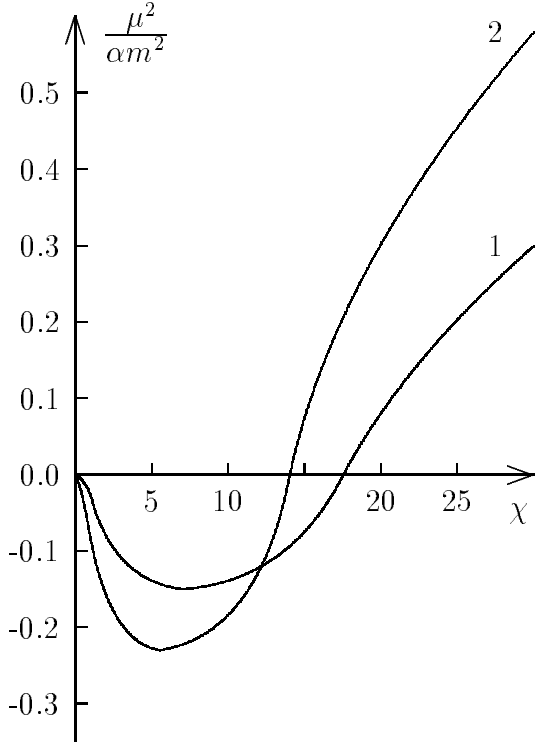}}
\caption{The photon dispersion curves in the crossed field.}
\label{fig:disp}
\end{figure}

We note that in a homogeneous electromagnetic field the energy-momentum
conservation law for axion-photon processes coincides formally with
the vacuum one~(\ref{eq:S0}). However, 
in calculating the probabilities of the processes one has to integrate
over the phase space of the final particles taking into account their
nontrivial field-induced kinematics. This is due to the fact that
an external field plays the role of a peculiar medium with
dispersion and absorption. In our case the external crossed field plays
the role of a homogeneous anisotropic medium.
The photon ``effective masses'' squared $\mu^2_\lambda$ induced by
the external field are defined as
the eigenvalues of the photon polarization operator:
\begin{equation}
\Pi_{\mu \nu} = i \; \sum_{\lambda = 1}^3 \mu_\lambda^2 \,
\frac{b^{(\lambda)}_\mu b^{(\lambda)}_\nu}{b^{(\lambda) 2}} , \;\;\;\;
b_{\alpha}^{(\lambda)} b_{\alpha}^{(\lambda')} =
\delta_{\lambda \lambda'} \;
(b_{\alpha}^{(\lambda)})^2 .
\label{eq:polop}
\end{equation}
Here $b^{(\lambda)}_\mu$ are the polarization operator eigenvectors
described in Appendix~\ref{app:basis}. We stress that only two eigenmodes
of the photon propagation with polarization vectors
\begin{eqnarray}
\varepsilon_\alpha^{(1)} & = &
\frac{b^{(1)}_\alpha}{\sqrt{- b^{(1) 2}}} =
\frac{ (q F)_{\alpha} }{ \sqrt{
( q F F q ) } },
\nonumber \\
\varepsilon _{\alpha}^{(2)} & = &
\frac{b^{(2)}_\alpha}{\sqrt{- b^{(2) 2}}} =
\frac{ (q \tilde F)_{\alpha} }
{ \sqrt{ (q \tilde F \tilde F q) } }
\label{eq:E1}
\end{eqnarray}
are realized in an external electromagnetic field.
The analysis of the photon polarization operator $\Pi_{\mu \nu}$
in a one-loop approximation in the crossed field~\cite{Ritus1}
shows that the dispersion curves corresponding to the
above-mentioned photon eigenmodes (Fig.~\ref{fig:disp}), though
being alike in their qualitative behavior, are different quantitatively.
%
%
%
%
It is seen from Fig.~\ref{fig:disp} that each dispersion curve has 
a negative minimum and changes its sign with increasing the dynamic 
parameter $\chi$.

The fact that both of the photon eigenmodes have negative ``squared 
masses'' in the region of $\chi \alt 15 - 18$ means that two-photon
decay is kinematically opened even for massless pseudoscalars
(e.g., arions~\cite{AU}). The difference in values of the
field-induced  ``squared masses'' of the first and second photon
eigenmodes makes possible the process of the photon
splitting $\gamma \to \gamma a$, where $a$ is an arbitrary, 
relatively light pseudoscalar.

In general, it is necessary to analyze the influence of the external
field on the axion mass $m_a$. The field-induced contribution to $m_a$
is connected with the real part of the amplitude
$\Delta M$ of $a \to f \tilde f \to a$
transition via the fermion loop by the relation:
\begin{eqnarray}
\delta m_a^2 = - \mbox{Re} \, \Delta M.
\label{eq:axmass}
\end{eqnarray}
To obtain a correct result for $\Delta M$ one has to
use the Lagrangian with the derivative as it was first emphasized
by Raffelt and Seckel~\cite{Raffelt88}:
\begin{equation}
{\cal L}_{af} =  \frac{ g_{af}}{2 m_f}
(\bar f \gamma_\mu \gamma_5 f) \, \partial_\mu \, a.
\label{eq:Ld1}
\end{equation}

%
\begin{figure}[tb]
\centerline{\epsfxsize=.4\textwidth \epsffile[185 410 435 520]{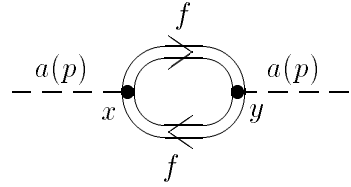}}
\caption{The diagram of the axion transition $a \to a$
         in an external electromagnetic field.}
\label{fig:polopaxion}
\end{figure}
%
%
In the second order of the perturbation theory the amplitude
$a \to f \tilde f \to a $ is described by the diagram in
Fig.~\ref{fig:polopaxion} and can be presented in the form:
\begin{eqnarray}
\Delta M = & - & i \, \frac{g^2_{af}}{4 m_f^2} \,
\int d ^4 Z \, e^{-i (q Z)}
\label{eq:P1} \\
& \times &
\Sp \, [ S(-Z) \, (q \gamma) \, \gamma_5 \, S(Z) \,
(q \gamma) \,\gamma_5 ] ,
\nonumber
\end{eqnarray}
where $Z = x - y$. The integration with respect to the
four-coordinate $Z$ is reduced to the generalized Gaussian integrals
of the type~(\ref{eq:GI}) from Appendix~\ref{app:vertex}
and results in 
\begin{eqnarray}
\Delta M & = &  \frac{g^2_{af}}{16 \pi^2}
 \bigg \lbrace
2 q^2 \int\limits_0^{\pi/2} \sin \phi \, d\phi \,
\int\limits_0^\infty \, \left (e^{-i \Phi} - e^{-i \Phi_0} \right ) \,
{dt \over t}
\nonumber \\
& - & m_f^2 \chi^2 \int\limits_0^{\pi/2} \sin^3 \phi \, d\phi \,
 \int\limits_0^\infty e^{-i \Phi} t \; dt
 \bigg \rbrace,
\label{eq:P0} \\
\Phi & = & \Phi_0 + {t^3 \chi^2 \sin^4 \phi \over 48} , \quad
\Phi_0   =  t \;\left ( 1 - \frac{q^2 \sin^2 \phi}{4 m_f^2} \right ) .
\nonumber
\end{eqnarray}
The main contribution to $\delta m_a^2$ has the following 
asymptotic behavior at large values of the dynamic parameter: 
%
\begin{equation}
\delta m_a^2 \simeq \frac{g_{af}^2 m_f^2}{24 \pi^{3/2}} \,
  \Gamma^2 (2/3) \, (6 \chi)^{2/3} ,
\label{eq:axmass-as}
\end{equation}
and permits one to estimate the field-induced correction 
to the axion mass:
\begin{equation}
\frac{\delta m_a^2}{m^2_a} \simeq 10^{- 9} 
\left ( {C_f m_f^2 \over 1 \, \mbox{MeV}^2} \right )^2 \chi^{2/3}.
\label{eq:MA}
\end{equation}
Here we used that $m_a$ and $g_{af}$ are connected with the
PQ symmetry-breaking scale $f_a$ by virtue of the relationships
$m_a = 0.62 \, \mbox{eV} \times (10^7 \, \mbox{GeV} /f_a)$ and
$g_{af} = C_f m_f / f_a$~\cite{Raffelt-book}.
Equation~(\ref{eq:MA}) shows that we can neglect the influence of an
external electromagnetic field on $m_a$.

\section{The axion decay $\lowercase{a} \to \gamma + \gamma$}

\subsection{The decay amplitude}

The expression for the effective
$a\gamma\gamma$ interaction~(\ref{eq:V1})
is used to study the ultrarelativistic ($E_a \gg m_a$)
axion decay $a(p) \to \gamma (q_1) + \gamma(q_2)$.
Because of the smallness of the axion mass ($m_a \alt 10^{-2}$~eV),
even axions with the energy of several eV are ultrarelativistic.
Assuming that the field-induced photon ``masses'' are also small in
comparison with their energies, below we will consider photons as
massless. This case corresponds to the collinear kinematics:
\begin{displaymath}
p_\mu \sim q_{1\mu} \sim q_{2\mu}.
\end{displaymath}

Before starting to study the physically most interesting case of
the ultrarelativistic axion decay let us analyze all tensor
structures of the effective axion-photon vertex~(\ref{eq:V1}), 
which define the decay amplitude, in an attempt 
to reveal the dominating contribution. Note that for real photons
the terms with $J_3$ and $\tilde J_3$ are suppressed by the smallness 
of the field-induced photon ``masses''
$(q_i f_i)_\alpha \sim \mu^2 \ll E_a^2 $.
It is natural to denote the other terms in Eq.~(\ref{eq:V1}) 
as $M^{(0)}$, $M^{(1)}$, and $M^{(2)}$ depending on the power 
of the external field tensor. 

Due to the Lorentz invariance the amplitude can be analyzed in any frame.
We will analyze it in the rest frame of the decaying axion with the
components of the four-momenta $p$, $q_1$, and $q_2$ being of order $m_a$.
In this case it is sufficient to allow for the order
of the dimensional quantities $M^{(0)}$, $M^{(1)}$, and $M^{(2)}$:
\begin{eqnarray}
M^{(0)} & \sim & \frac{1}{m_f} (f_1 \tilde f_2) \sim \frac{m_a^2}{m_f},
\nonumber \\
M^{(1)} & \sim & \frac{1}{m_f^3} (f_1 \tilde F f_2)
\sim \frac {m_a^2}{m_f^3} \; F',
\label{eq:Mrest}\\
M^{(2)} & \sim & \frac{1}{m_f^5} (f_1 \tilde F) (f_2 F)
\sim \frac {m_a^2}{m_f^5} \; {F'}^2,
\nonumber
\end{eqnarray}
where $F'$ stands for the strengths of the magnetic and electric fields
in the axion rest frame. In this case the electromagnetic field
in Eq.~(\ref{eq:Mrest}) is obtained by the Lorentz transformation from
the laboratory frame, in which the external field $F$ is given,
to the rest frame of the decaying axion:
\begin{equation}
F' \sim {E_a \over m_a} F \gg  F.
\label{eq:UF}
\end{equation}
In view of Eq.~(\ref{eq:UF}), the expressions~(\ref{eq:Mrest})
can be written in the form 
\begin{eqnarray}
M^{(0)} & \sim & \frac{m_a^2}{m_f},
\nonumber \\
M^{(1)} & \sim & \frac{m_a E_a}{m_f^3} \; F,
\label{eq:Multra}\\
M^{(2)} & \sim & \frac{E_a^2}{m_f^5} \; F^2.
\nonumber
\end{eqnarray}
The following is seen from Eqs.~(\ref{eq:Multra}). 
\begin{enumerate} 
\item
The external electromagnetic field affects substantially
the ultrarelativistic axion decay because the field reduces the
suppression caused by the smallness of the axion mass.
A similar catalyzing effect of the external electromagnetic fields
of various configurations on the radiative decay of the massive
neutrino was discovered in~\cite{Ljuba92-94,Ljuba96}.
\item
The contribution to the amplitude bilinear in the external field
dominates because it does not contain the above-mentioned
suppression factor.
\end{enumerate}
With regard to the analysis performed only terms bilinear in the 
external field should be kept in the amplitude of the ultrarelativistic 
axion decay:
\begin{eqnarray}
M & \simeq & \frac{\alpha}{\pi} \sum_f \frac{Q^2_f g_{af}}{m_f}
\label{eq:M1} \\
& \times &
\left \{ \frac{\chi}{\chi_2} (f_1 \tilde {\cal F}) (f_2 {\cal F}) J
+ ( \varepsilon_1, q_1 \leftrightarrow \varepsilon_2, q_2 )
\right \} ,
\nonumber
\end{eqnarray}

\begin{eqnarray}
J(\chi_1, \chi_2) & = & \int\limits_0^1 dx \int \limits_0^{1-x}
dy \, y \, (1 - y - 2x) \, \eta^3 \; (1 - \eta \, f(\eta)),
\nonumber \\
\eta & = & \Big \{ \left [ x (1 - x) \chi_2 - y (1 - y) \chi_1 \right ]^2
\label{eq:IntJ} \\
     & + & 4 x y \left [ x y + (1 - x - y )(x + y) \right ] \chi_1 \chi_2
\Big \}^{-1/3},
\nonumber
\end{eqnarray}
where $f(\eta)$ is the Hardy-Stokes function defined in Eq.~(\ref{eq:MT}),
$\chi$, $\chi_1$, and $\chi_2$ are the dynamic parameters determined
in Eq.~(\ref{eq:V2}). We have neglected terms of order $m_a^2/E_a^2$,
$\mu_i^2/E_a^2$ in the argument of the function $f(\eta)$.

\subsection{The decay probability}

\begin{figure}[tb]
\centerline{\epsfxsize=.4\textwidth \epsffile[95 315 535 600]{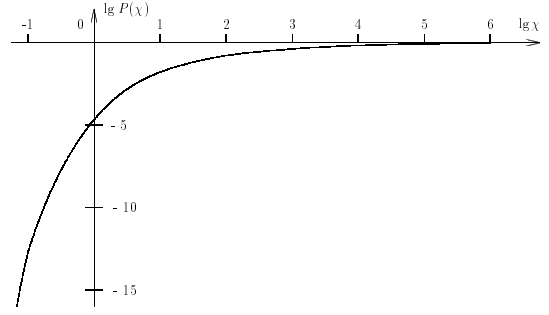}}
\caption{The plot of the decay factor $P (\chi)$
         as a function of the dynamic parameter $\chi$.}
\label{fig:graph}
\end{figure}

The probability of the field-induced decay of the axion
has the following form:
\begin{eqnarray}
W^{(F)} & = & {1 \over 2!} \frac{1}{32 \pi^2 E_a}
\int \! \frac{d^3 q_1 d^3 q_2}{\omega_1 \omega_2} \,
\delta^{(4)} ( p - q_1 - q_2 ) \sum_{\lambda,\lambda'}
\vert M \vert^2
\nonumber \\
& = &
{1 \over 32 \pi E_a} \int \limits_{0}^1 d t \;
\sum_{\lambda,\lambda'} \vert M \vert^2.
\label{eq:WFdef}
\end{eqnarray}
The factor $1/2!$ in the integration over the final
state takes into account the identity of two final photons.
The process considered is two-body decay, but its amplitude is
not a constant as in vacuum because it depends on the external
electromagnetic field tensor. That is why the calculation of the decay
probability is not reduced to multiplying the matrix element squared
with the final phase space. Under the summation of the polarizations
$\lambda$, $\lambda'$ in Eq.~(\ref{eq:WFdef}) it is convenient to use the
polarization vectors~(\ref{eq:E1}). The expression for the probability is 
\begin{eqnarray}
W^{(F)} & = & {1 \over \pi E_a}
\left( {\alpha \over \pi} \right )^2
\int \limits_{0}^1 d t \; t^2 \;
\label{eq:WF1} \\
& \times &
\left \vert \sum \limits_f
Q^2_f g_{af} m_f \chi^2 \; J(t \chi, (1 - t) \chi)
\right \vert ^2.
\nonumber
\end{eqnarray}
With the strong hierarchy of the fermion masses, 
\begin{eqnarray}
{\chi_{f_1}^2 \over \chi_{f_2}^2 } = \left (
m_{f_2} \over m_{f_1} \right )^6  \gg 1, 
\nonumber
\end{eqnarray}
the contribution from the fermion with the maximum value
of the dynamic parameter $\chi^{2} = e_f^2 (p F F p)/m_{f}^{6}$
can dominate in Eq.~(\ref{eq:WF1}):
\begin{eqnarray}
W^{(F)} \simeq  3.32 \left( \frac{\alpha}{\pi} \right)^2
{(Q_f^2 g_{af} m_f)^2 \over \pi E_a} P (\chi).
\label{eq:RP}
\end{eqnarray}
The plot of $P(\chi)$ is shown in Fig.~\ref{fig:graph}.
%
%
%
%
We also give the asymptotic behavior of the function $P(\chi)$
for both small and large values of the dynamic parameter $\chi$:
\begin{eqnarray}
{P(\chi) \bigg \vert_{\chi \ll 1} } & \simeq & 2.31 \times 10^{-5} 
\chi^8 ( 1 + 9.5\chi^2 + \cdots ),
\nonumber \\
{P(\chi) \bigg \vert _{\chi^{1/3} \gg 1} } & \simeq &
1 - {8.8 \over \chi^{1/3}} + {32.0 \over \chi^{2/3}}
- \cdots .
\nonumber
\end{eqnarray}
As it is seen from the asymptotic behavior of the function
$P(\chi)$ at large $\chi$, the decay probability~(\ref{eq:RP})
does not depend on the dynamic parameter.
Note that a similar behavior of the amplitude and probability
at $\chi \gg 1$ takes place in the photon splitting
$\gamma \to \gamma \gamma$~\cite{Ritus}.

To illustrate the catalyzing influence of the external field
on the ultrarelativistic axion decay, let us compare Eq.~(\ref{eq:RP})
with the well-known axion decay probability in
vacuum~\cite{Raffelt90,Raffelt-book}:
\begin{eqnarray}
W_0 = \frac {g_{a\gamma}^2 m_a^4 }{64 \pi E_a}.
\label{eq:W0}
\end{eqnarray}
Here $g_{a \gamma} = (\alpha / 2 \pi f_a) \times 
(E/N - 1.92 \pm 0.08)$~\cite{Raffelt-book},
where $E$ and $N$ are model-dependent coefficients of the
electromagnetic and color anomalies. A comparison of
Eqs.~(\ref{eq:RP}) and~(\ref{eq:W0}), 
\begin{equation}
R_f = {W^{(F)} \over W_0} \simeq  2.12 \times 10^2
\left ( \frac{\alpha Q_f g_{af} m_f}{\pi g_{a \gamma} m^2_a} \right )^2 
P(\chi) , 
\label{eq:R1}
\end{equation}
demonstrates a strong catalyzing influence of
the external field on the ultrarelativistic axion decay
($E_a \gg m_a$), because $W^{(F)}$ has no
suppression factor associated with the smallness of the axion mass,
except the coupling constants ($g_{af}, g_{a\gamma} \sim f_a^{-1}$).
Let us estimate in Eq.~(\ref{eq:R1}) the contribution from the electron 
in the loop because this fermion is the most sensitive to the external 
field:
\begin{eqnarray}
R_e \simeq 10^{37}
\left ( {\cos^2\beta \over E/N - 1.92} \right )^2 \;
\left ( {10^{-3} \mbox{eV} \over m_a } \right )^4 \; P(\chi).
\label{eq:R2}
\end{eqnarray}
Here $\cos^2 \beta$ determines the electron Yukawa coupling
$g_{ae} = \frac {1}{3}\cos^2\beta\; (m_e/f_a)$ in the DFSZ 
model~\cite{DFSZ}. Large values of the dynamic parameter 
$\chi \gg 1$, when $P(\chi) \sim 1$,
can be realized, for example, in the case of the decaying
axion energy $\sim 10$~MeV and the magnetic-field strength
$B \sim B_e \simeq 4.41 \times 10^{13}$~G.

\section{Photon splitting $\gamma \to \gamma + \lowercase{a}$}

As it was pointed in Sec.~III the external field has a substantial
influence on the photon dispersion relation so that the photon
splitting $\gamma \to \gamma a$ is allowed kinematically due to the
difference in the field-induced effective photon ``masses''.
This process can play a role as a possible additional mechanism of
energy loss by astrophysical objects. As this process is a crossing
channel of the axion decay $a(p) \to \gamma(q_1) + \gamma(q_2)$ and
its amplitude can be easily obtained from the amplitude~(\ref{eq:M1})
using the replacement:
\begin{equation}
q_1 \to - q_1, \quad p \to - p,
\end{equation}
which corresponds to
\begin{equation}
f_1 \to - f_1, \quad \chi \to - \chi, \quad \chi_1 \to - \chi_1.
\end{equation}
The kinematics of this process is close to the collinear one
as well as in the ultrarelativistic axion decay. Note that in studying
this process in different ranges of the dynamic parameter one should take
into account the specific dispersion behavior of the photon modes in the
initial and final states.

\subsection{The limit of small values of the dynamic
            parameter $\chi_1 \ll 1$}

In the case of small values of the initial photon dynamic parameter
$\chi_1$ the splitting of the photon with the first
polarization~\footnote{Numbering of the photon types is made by their
polarizations in accordance with Eq.~(\ref{eq:E1}).} 
is allowed kinematically only due to the condition 
$\mu^2_1 > \mu^2_2$ (see Fig.~\ref{fig:disp}):
\begin{eqnarray}
\gamma^{(1)} \rightarrow \gamma^{(2)} + a.
\label{eq:G1}
\end{eqnarray}
With the photon polarization vectors~(\ref{eq:E1}) the
amplitude can be presented in the form:
\begin{eqnarray}
 M  & \simeq & - \frac{4 \alpha}{\pi} t\;(1 - t) \sum_f \;
Q^2_f g_{af} m_f \; \chi_1^2 \; J(t \chi_1, \chi_1) ,
\label{eq:M5}
\end{eqnarray}
where $t = \omega_2/\omega_1$ is the relative energy of the final photon. 
The function $J$ defined in Eq.~(\ref{eq:IntJ}) has the following 
asymptotic behavior at small values of its arguments:
\begin{eqnarray}
J (t \chi_1, \chi_1) \bigg \vert_{\chi_1 \ll 1}  \simeq \frac{2}{63} \;
\chi_1^2 \; (1 - 2t) + O(\chi_1^4).
\nonumber
\end{eqnarray}

The probability of the photon splitting is 
\begin{eqnarray}
W^{(F)} & = & {1 \over 16\pi \omega_1}
\int \limits_{0}^1 d t \, \vert  M \vert^2
\label{eq:WFll1} \\
& \simeq &
4.8 \times 10^{-6} \,
\left( {\alpha \over \pi}\right )^2
\frac {\left( \sum_f Q^2_f g_{af} m_f \chi_1^4  \right )^2}
{\pi \omega_1}.
\nonumber
\end{eqnarray}
  
To illustrate a possible application of the result obtained
we estimate the contribution of this process to the axion
emissivity $Q_a$ of the photon gas:
\begin{eqnarray}
Q_a & = & \int \; \frac{d^3 q_1}{(2 \pi)^3} \;
\omega_1 \, n_B (\omega_1) \;
\label{eq:Q1} \\
& \times &
\int \limits_0^1 d t \, \frac{d W^{(F)}}{d t}\; (1 - t) \;
[1 + n_B (\omega_1 t)],
\nonumber
\end{eqnarray}
where $n_B(\omega_1)$ and $n_B(\omega_1 t)$ are the Planck distribution
functions of the initial and final photons at temperature $T$, respectively.
In Eq.~(\ref{eq:Q1}) we have taken into account that the photon
of only one polarization (the first one in this case) splits.
Note that the dynamic parameter $\chi_1$ in $d W^{(F)}/d t$
depends on the photon energy $\omega_1$ and the angle $\theta$ between
the initial photon's momentum  ${\bf q_1}$ and the magnetic-field
strength ${\bf B}$:
\begin{eqnarray}
\chi_1 = \frac{\omega_1}{m_f}\; \frac{B}{B_f} \; \sin \theta,
\nonumber
\end{eqnarray}
where $B_f = m_f^2 / e Q_f$ is the critical value 
of the magnetic-field strength for the given fermion. 
Neglecting the photon ``effective masses'' squared
($\omega_1^2 = | {\bf q_1} |^2 + \mu_1^2 \simeq | {\bf q_1} |^2$),
the result of the calculation of the axion emissivity~(\ref{eq:Q1})
can be presented in the form 
\begin{eqnarray}
Q_a & \simeq & C \, \frac{\alpha^2}{\pi^5} \left (
\sum_f \frac{Q_f^2 \, g_{af}}{m_f^3 \, B_f^4} \right )^2 
T^{11} \, B^8 .
\label{eq:Q2}
\end{eqnarray}
The numerical factor $C$ is expressed in terms of an infinite sum 
of the Hurwitz $\zeta$~functions
$\zeta(k,z) = \sum \limits_{n=0}^\infty 1/(n + z)^k$:
\begin{eqnarray}
C & = & \frac{1}{63^2}\;
\int \limits_0^\pi \sin^9 \theta d \theta \;
\int \limits_0^1 d t\; t^2 \;
(1 - t)^3\; (1 - 2t)^2 \;
\nonumber \\
& \times &
\int \limits_0^\infty \frac{x^{10} d x}{(e^x - 1)(1 - e^{- xt})}
\nonumber \\
& = & \!
\frac{2^{13}}{63^2} \sum_{n = 1}^\infty \left (
\frac{ \zeta(3,n)}{n^8}
- \frac{4 \zeta(5,n)}{7 n^6}
+ \frac{3 \zeta(7,n)}{7 n^4}
\right ) = 2.15145 ,
\nonumber
\end{eqnarray}
where $x = \omega_1/T$ is the relative energy of the decaying photon.

Let us estimate the contribution of the photon splitting
$\gamma \to \gamma a$ into the axion luminosity in a supernova
explosion from a region of order of a hundred kilometers in size
outside the neutrinosphere. In this region a rather rarefied plasma
with the temperature of order of MeV and the magnetic field of order
$10^{13}$~G can exist. Under these conditions the axion luminosity is 
\begin{eqnarray}
L_a & \simeq & 3 \times 10^{36} \; \frac{\mbox{erg}}{\mbox{s}}
\left ( \frac{g_{ae}}{10^{-13}} \right )^2
\left ( \frac{T}{1 \, \mbox{MeV}} \right )^{11}
\label{eq:Lum} \\
& \times &
\left ( \frac{B}{10^{13} \, \mbox{G}} \right )^8
\left ( \frac{R}{10^3 \, \mbox{km}}\right )^3 .
\nonumber
\end{eqnarray}
The comparison of Eq.~(\ref{eq:Lum}) with the total neutrino luminosity
$L_\nu \sim 10^{52}$~erg/s from the neutrinosphere shows that the
contribution of the photon splitting to the energy loss is negligibly
small. Here we used the strongest restriction on the axion coupling
constant with electrons is
$g_{ae} \sim 10^{-13}$~\cite{Raffelt-castle97,Raffelt-book,Altherr}.

\subsection{The limit of large values of the dynamic
            parameter  $\chi_1 \gg 1$}

At large values of the dynamic parameter $\chi_1$ the field-induced
``effective mass'' squared of the photon of the second polarization
becomes greater than that of the photon of the first polarization
($\mu^2_2 > \mu^2_1$, see Fig.~\ref{fig:disp}).
In this case the channel
\begin{eqnarray}
\gamma^{(2)} \rightarrow \gamma^{(1)} + a
\label{eq:G2}
\end{eqnarray}
is kinematically opened only.
The process amplitude at large $\chi_1$ is 
\begin{eqnarray}
 M  & \simeq & \frac{4 \alpha}{\pi} \;(1 - t) \sum_f \;
Q^2_f g_{af} m_f \; I (t) ,
\label{eq:M6} \\
I (t) & = & \int \limits_{D} \; \frac{y \, (1 - y - 2 x) \, dx dy}
{[ x (1 - x) t - y (1 - y) ]^2 + 4 x y t \, (1 - x - y)^2},
\nonumber
\end{eqnarray}
where the integration is carried out over the area $D$: $x, y \ge 0$,
$x + y \le 1$. It is seen from Eq.~(\ref{eq:M6}) that the amplitude does
not depend on the dynamic parameter. 

With the amplitude~(\ref{eq:M6}) the splitting probability has the form:
\begin{eqnarray}
W^{(F)} & \simeq & 2.5 \,
\left( {\alpha \over \pi}\right )^2
\frac {\left( \sum_f Q^2_f g_{af} m_f  \right )^2}
{\pi \omega_1}.
\label{eq:WFgg1}
\end{eqnarray}

In real astrophysical objects with strong magnetic fields
the temperature $T \agt 10$~MeV can exist in the central regions where
from both components of the active medium, the magnetic field and a plasma,
the plasma component substantially dominates. Thus, the expressions
obtained are not applied for the estimation of the luminosity
of astrophysical objects because we have taken into account the
influence of the external field only.

\section{Conclusion}

In this work we have investigated the effective
$a \gamma \gamma$ in\-te\-rac\-tion
($a$ is a pseudoscalar particle) induced by an external constant 
crossed field. Calculations of processes in this electromagnetic 
field configuration are very general, being a 
relativistic limit of calculations in an arbitrary weak smooth field. 
For the pseudoscalar particle we considered the axion, 
the most widely discussed particle corresponding to the spontaneous
breaking of the PQ symmetry.

In the third order of the perturbation theory, the field-induced
vertex $a \gamma\gamma$~(\ref{eq:V1}) was obtained
for the case when all the particles were off the mass shell.
This vertex can be used as the effective Lagrangian of the
$a \gamma \gamma$ interaction in the investigation of processes
involving axion or other pseudoscalar particles with a coupling
of the type~(\ref{eq:L1}).

As a specific example of the way this vertex can be used we considered
the axion decay $a \rightarrow \gamma + \gamma$ in an external field.
The dominant contribution to this process (not suppressed by the axion mass)
is presented in Eq.~(\ref{eq:M1}).
The decay probability~(\ref{eq:RP}) is calculated in the limits of both
large and small values of the dynamical parameter $\chi$.
Comparison of Eq.~(\ref{eq:RP}) with the probability of the two-photon
axion decay in vacuum shows a strong effect of the external field
($\sim 10^{37}$) on the decay of the ultrarelativistic axion
($E_a \gg m_a$).

As another example, we have studied the photon splitting 
$\gamma \to \gamma + a$ which could be of interest as an additional
mechanism of energy losses by astrophysical objects.
This forbidden in vacuum process becomes kinematically possible
because photons of different polarizations obtain different
field-induced effective ``squared masses''.
At the same time an external field influence on the axion mass is
negligible. The axion luminosity estimated for supernova conditions
turn out to be small in comparison with the neutrino luminosity.

\acknowledgments

The authors are grateful to G.G.~Raffelt and V.A.~Ru\-ba\-kov for
interest in this paper and useful critical remarks. This work was
partially supported by the INTAS under Grant No.~96-0659 and by the
Russian Foundation for Basic Research under Grant No.~98-02-16694.
The work of N.V.~Mikheev was supported under Grant No.~d98-181
by the International Soros Science Education Program.
N.M. and L.V. acknowledge the support by CERN during a visit when 
this paper was partly done. 

\appendix


\section{The propagator of a charged fermion in a crossed field}
\label{app:propagator}

A charged fermion propagator $S^{(F)} (x, y)$ in the external
crossed field in the proper time formalism~\cite{Schwinger,Itzykson}
can be presented in the relativistically invariant form~\cite{Ljuba96} 
\begin{eqnarray}
S^{(F)} (x,y) & = & e^{\mbox{\normalsize $i \Omega (x,y)$}} S (X),
\label{eq:PSV} \\
\Omega(x,y) & = & - e_f \, \int_y^x d\xi^\mu \, \left [ A_\mu (\xi) +
            \frac{1}{2} F_{\mu \nu} (\xi - y)^\nu \right ] ,
\nonumber \\
S (X) & = & - {i \over 16 \pi^2} \int\limits_0^\infty
{ds \over s^2} \bigg [ {1 \over 2s} (X \gamma) -
{i e_f \over 2} (X \tilde F \gamma) \gamma_5
\nonumber \\
& - & {s e_f^2 \over 3} (X F F \gamma) + m_f
+ {s m_f e_f \over 2} (\gamma F \gamma) \bigg ]
\nonumber \\
& \times & \exp \left [ -i \left ( s m_f^2 + {X^2 \over 4 s} +
{s e_f^2 \over 12} (XFFX) \right ) \right ] ,
\nonumber
\end{eqnarray}
where $X_\mu = (x - y)_\mu$; $m_f$ and $e_f$ are the mass and charge
of the virtual fermion ($e_f = e Q_f$, $e > 0$ is the elementary charge,
$Q_f$ is the relative fermion charge);
$A_\mu$, $F_{\mu \nu}$, and $\tilde F_{\mu \nu}$ are the four-potential,
tensor, and dual tensor of the crossed field;
$(X \gamma) = X_\mu \gamma^\mu$,
$\gamma^\mu$, $\gamma_5$ are the Dirac $\gamma$ matrices.

The inclusion of translationally noninvariant factors $\Omega(x_i,x_{i+1})$
arising from the corresponding propagators $S^{(F)}(x_i,x_{i+1})$
in calculations of $n$-point loop diagrams is discussed in detail
in Appendix~\ref{app:bable}.

\section{Inclusion of translationally noninvariant factors
         $\Omega(\lowercase{x,y})$ of fermion propagators
         in $\lowercase{n}$-point loop diagram}
\label{app:bable}

In this Appendix we discuss the inclusion of the translationally
noninvariant factor $\Omega(x,y)$ from the propagator $S^{(F)} (x,y)$ 
Eq.~(\ref{eq:PSV}) in the calculation of $n$-point loop diagrams. 
In particular, the diagrams in Fig.~\ref{fig:diag1} describing the 
effective $a \gamma \gamma$ interaction are three-point loop diagrams.

\begin{figure}
\centerline{\epsfxsize=.4\textwidth \epsffile[230 390 415 480]{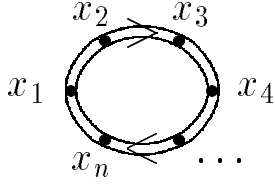}}
\caption{$n$-point loop fermion diagram with cut legs
         in an external electromagnetic field.}
\label{fig:loop}
\end{figure}

The four-potential $A_\mu (x)$ of a constant external electromagnetic 
field can be written as 
\begin{eqnarray}
A_\mu = \frac{1}{2} (x F)_\mu = \frac{1}{2} x^\nu F_{\nu\mu}.
\label{eq:Poten}
\end{eqnarray}
With Eq.~(\ref{eq:Poten}), the expression for $\Omega (x_1, x_2)$
is substantially simplified:
\begin{eqnarray}
\Omega (x_1, x_2) & = & - e_f \int \limits_{x_2}^{x_1} d \xi^\mu
\left [ A_\mu (\xi) + \frac{1}{2} F_{\mu \nu} (\xi - x_2)^\nu \right ]
\nonumber \\
& = & \frac{e_f}{2} (x_1 F x_2).
\label{eq:Om1}
\end{eqnarray}
The contribution from the translationally noninvariant factors
$\Omega(x_i, x_{i+1})$ Eq.~(\ref{eq:Om1}) to the amplitude of an $n$-point
diagram in Fig.~\ref{fig:loop}
%
%
%
%
is generally given by 
\begin{eqnarray}
\Omega_{\rm tot} & = & \frac{e_f}{2} \sum \limits_{i=1}^n (x_i F x_{i+1})
\bigg \vert_{ x_{n+1} = x_1 } = \frac{e_f}{4}
\sum \limits_{i=1}^{n-1} (Z_i F Z_{i+1}),
\nonumber \\
Z_i & = & x_i - x_{i+1},\; \; \; \; Z_n = x_n - x_1, \;\;\;\;
\sum \limits_{i=1}^{n} Z_i = 0.
\label{eq:Om2}
\end{eqnarray}
We note that, while for each propagator 
$S^{(F)} (x_i, x_{i+1})$ the phase $\Omega(x_i, x_{i+1})$ 
is generally translationally and gauge noninvariant, the total phase 
of the $n$-point fer\-mi\-on loop $\Omega_{\rm tot}$ Eq.~(\ref{eq:Om2})
only depends on the difference of the coordinates and satisfies
the requirement of gauge and translational invariance.

The simplest situation is realized for the two-point loop diagram
when the total phase is zero:
\begin{eqnarray}
\Omega_{\rm tot} & = & \Omega(x_1, x_2) + \Omega(x_2, x_1) =
- \frac{e_f}{4} (X F X) = 0,
\nonumber \\
X & = & x_1 - x_2.
\nonumber
\end{eqnarray}
For the three-point loop diagram studied in this paper we have 
\begin{eqnarray}
\Omega_{tot} & = &
\Omega(x_1, x_3) + \Omega(x_3, x_2) + \Omega(x_2, x_1) =
- \frac{e_f}{2} (X F Y),
\nonumber \\
X & = & x_1 - x_2, \;\;\; Y = x_1 - x_3.
\nonumber
\end{eqnarray}

\section{Calculation of the vertex function $\Lambda$}
\label{app:vertex}

In this Appendix we give the basic stages of the calculation of the
three-point vertex function $\Lambda$ describing the effective
$a \gamma \gamma$ interaction ($a$ is an arbitrary pseudoscalar
particle) in the external crossed field. After integration with respect
to the coordinate $x_1$ in Eq.~(\ref{eq:S1}) and separation of the
delta-function $\delta^{(4)} (p - q_1 - q_2)$, the expression for
$\Lambda$ can be represented according to Eq.~(\ref{eq:S0}) as 
\begin{eqnarray}
\Lambda & = & - i e_f^2 g_{af}
\int d^4 X \, d^4 Y \,
\label{eq:V3} \\
& \times &
 \exp \left ( - i \left \lbrace (q_1 X) + (q_2 Y) +
 \frac{e_f}{2} (X F Y) \right \rbrace \right )
\nonumber \\
& \times &
\mbox {Sp} \lbrace S(Y) (\varepsilon_2 \gamma)
S(X - Y) (\varepsilon_1 \gamma ) S(-X) \gamma_5 \rbrace
\nonumber \\
& + &
( \varepsilon_1, q_1 \leftrightarrow \varepsilon_2, q_2 ) .
\nonumber
\end{eqnarray}
With the explicit expression for the propagators~(\ref{eq:PSV})
the vertex function is 
\begin{eqnarray}
\Lambda & = & {\alpha \over \pi} \sum
\frac{Q_f^2 g_{af} m_f}{8 (4 \pi)^4}
\int \limits_0^{\infty} \frac{ ds dv d\tau }{ s^2 v^2 \tau^2 } \,
\mbox{$e$}^{\mbox{$-i m_f^2 (s + v + \tau)$}}
\nonumber \\
 & \times &
\int d^4 X d^4 Y \,
\bigg \lbrace ((X - Y) R_1 X) + (Y R_2 (Y - X))
\nonumber \\
& + &
(Y R_3 X) - {m_f^2 \over 4} R_0 \bigg \rbrace
\label{eq:Lam1} \\
 & \times &
\exp \bigg ( -i \bigg \lbrace (q_1 X) + (q_2 Y) +
{X^2 \over 4}  \bigg ( {1\over v} + {1\over \tau} \bigg )
\nonumber \\
& + &
{Y^2 \over 4} \bigg ( {1\over s} + {1\over v} \bigg )
- \frac{(XY)}{2v}
+ \frac{e_f^2}{12} (v + \tau) \, (XFFX) +
\nonumber \\
& + &
\frac{e_f^2}{12} (s + v) \, (YFFY) -
\frac{e_f^2 v}{6} (XFFY) - \frac{e_f}{2} (XFY)
\bigg \rbrace \bigg )
\nonumber \\
 & + &
 (\varepsilon_1, q_1 \leftrightarrow \varepsilon_2, q_2),
\nonumber
\end{eqnarray}
where $s$, $v$ and $\tau$ are proper time variables defined
by three propagators of the form~(\ref{eq:PSV}), $X = x_1 - x_2$,
$Y = x_1 - x_3$. In Eq.~(\ref{eq:Lam1}) all the dependence
on four-coordinates $X$ and $Y$ is shown explicitly in the
preexponential factor and in the exponent, so that~$R_0$ and
the tensors $(R_i)_{\alpha\beta}$ ($i = 1,2,3$) are functions 
of proper time variables and the constant external field tensor:
\begin{eqnarray}
R_0 & = & \Sp \lbrace \gamma_5 h_1(s)\,(\varepsilon_2 \gamma) \,
h_1(v)\,(\varepsilon_1 \gamma) \,h_1(\tau) \rbrace,
\nonumber \\
(R_1)_{\alpha\beta} & = & \Sp \lbrace \gamma_5
h_1(s)\,(\varepsilon_2 \gamma) \,
h_{2\alpha}(v)\,(\varepsilon_1 \gamma) \,
h_{2\beta}(\tau) \rbrace,
\nonumber \\
(R_2)_{\alpha\beta} & = & \Sp \lbrace \gamma_5 h_1(\tau)\,
h_{2\alpha}(s)\,(\varepsilon_2 \gamma) \,
h_{2\beta}(v)\,(\varepsilon_1 \gamma) \rbrace,
\nonumber \\
(R_3)_{\alpha\beta} & = & \Sp \lbrace \gamma_5
h_1(v)\,(\varepsilon_1 \gamma) \,
h_{2\beta}(\tau)\, h_{2\alpha}(s)\,
(\varepsilon_2 \gamma) \rbrace,
\nonumber \\
h_1(s) & = & 2 + e_f s (\gamma F \gamma),
\nonumber \\
h_{2\alpha}(s) & = & \frac{1}{2s}\, \gamma_\alpha -
\frac{e^2_f s}{3}\, (\gamma F F)_\alpha +
\frac{i e_f}{2}\, (\gamma \tilde F)_\alpha \gamma_5 .
\nonumber
\end{eqnarray}
The integration with respect to the four-coordinates~$X$ and~$Y$
in Eq.~(\ref{eq:Lam1}) reduces to the calculation of the generalized
Gaussian integrals of three types: scalar, vector, and
tensor of rank two. The calculation method of such integrals which 
arise in the integration of the two-point loop diagrams is given
in Ref.~\cite{Ljuba96}. However the integration with respect to the
coordinates is significantly simplified by formally introducing the
eight-coordinate~$Z$ and the eight-momentum~$Q$ defined in terms of 
the four-coordinates~$X$, $Y$ and the four-momenta of photons~$q_1$, 
$q_2$ in the following way:
\begin{equation}
Z_\mu = \left (
\begin{array}{c}
X_\mu \\ Y_\mu
\end{array}
\right ) ,
\qquad
Q_\mu = \left (
\begin{array}{c}
q_{1\mu} \\ q_{2\mu}
\end{array}
\right ) .
\label{eq:ZQ}
\end{equation}
In terms of the generalized coordinate~$Z_\mu$ and the generalized
momentum~$Q_\mu$ the vertex function~(\ref{eq:Lam1})
can be written in the most compact form:
\begin{eqnarray}
\Lambda & = & \frac{\alpha}{\pi} \sum_f \frac{Q^2_f g_{af} m_f}
{8 (4 \pi)^4} \int\limits_0^\infty
{ds dv d\tau \over s^2 v^2 \tau^2} \,
\mbox{$e$}^{\mbox{$- i m_f^2 (s + v + \tau)$}}
\nonumber \\
& \times & \int d^8Z \,
\mbox{$e$}^{\mbox{$- i \left [ \frac{1}{4} (Z G Z) + (QZ) \right ]$}}
\Big [ R_0 + (ZNZ) \Big ]
\nonumber \\
& + &
( \varepsilon_1, q_1 \leftrightarrow \varepsilon_2, q_2 ).
\label{eq:Lam2}
\end{eqnarray}
The matrices $G$ and $N$ can be presented as the Kronecker
product of ($2 \times 2$) matrices and Lorentz tensors of
rank two [($4 \times 4$)-matrices]: 
\begin{eqnarray}
G_{\alpha\beta} & = &
g_{\alpha\beta} \, G_1
+ e_f \, F_{\alpha\beta} \, G_2
+ e_f^2 \, (F F)_{\alpha\beta}\, G_3 ,
\label{eq:G} \\ 
G_1 & = & 
\frac{1}{s v \tau}
\left (
\begin{array}{cc}
  s (v + \tau) &  - s \tau \\
  - s \tau  & \tau (s + v)
\end{array}
\right ),
\nonumber \\
G_2 & = & 
\left (
\begin{array}{cc}
  0 & 1 \\
  - 1 & 0
\end{array}
\right ),
\qquad
G_3 =
\frac{1}{3}
\left (
\begin{array}{cc}
  v + \tau &  - v \\
  - v  & s + v
\end{array}
\right ),
\nonumber 
\end{eqnarray}

\begin{eqnarray}
N_{\alpha\beta} = N_1 (R_1)_{\alpha\beta} +
N_2 (R_2)_{\alpha\beta} + N_3 (R_3)_{\alpha\beta},
\nonumber  
\end{eqnarray}

\begin{eqnarray}
N_1 & = & 
\left (
\begin{array}{cc}
  1 & 0 \\
  - 1 & 0
\end{array}
\right ),
\,
N_2 =
\left (
\begin{array}{cc}
  0 & 0 \\
  - 1 & 1
\end{array}
\right ),
\,
N_3 =
\left (
\begin{array}{cc}
  0 & 0 \\
  1 & 0
\end{array}
\right ).
\nonumber
\end{eqnarray}
Thus, the integration with respect to the coordinate $Z_\mu$ in
Eq.~(\ref{eq:Lam2}) has been reduced to the calculation of the
generalized Gaussian integrals of two types~-- the scalar and 
tensor of rank two: 
\begin{eqnarray}
J & = & \int d^8Z \,
\mbox{$e$}^{\mbox{$- i \left [ \frac{1}{4} (ZGZ) + (QZ) \right ]$}}
\label{eq:GI} \\
& = & - (4 \pi )^4 \frac{s^2 v^2 \tau^2}{(s + v + \tau)^2} \,
\mbox{$e$}^{\mbox{$i (Q G^{-1} Q)$}},
\nonumber \\
T_{\mu \nu} & = & \int d^8Z \, Z_\mu Z_\nu \,
\mbox{$e$}^{\mbox{$- i \left [ {1 \over 4} (ZGZ) + (QZ) \right ]$}}
= - {\partial^2 J \over \partial Q^\mu \partial Q^\nu}
\nonumber \\
& = & \left [ 4 (Q G^{-1})_\mu \,(G^{-1} Q)_\nu -
2 i G^{-1}_{\mu\nu} \right ] \, J,
\nonumber
\end{eqnarray}
where $G^{-1}$ is the inverse of the matrix~(\ref{eq:G}):
\begin{equation}
G^{-1}_{\alpha\beta} = g_{\alpha\beta} \,\tilde G_1
+ e_f \, F_{\alpha\beta}\, \tilde G_2
+ e_f^2 \, (F F)_{\alpha\beta} \,\tilde G_3,
\label{eq:GT}
\end{equation}
\begin{displaymath}
\tilde G_1 =
\frac{1}{z}
\left (
\begin{array}{cc}
  \tau (s + v ) &   s \tau \\
   s \tau  & s (v + \tau)
\end{array}
\right ),
\,
\tilde G_2 =
\frac{s v \tau}{z}
\left (
\begin{array}{cc}
  0 & - 1 \\
  1 & 0
\end{array}
\right ),
\end{displaymath}
\begin{eqnarray}
(\tilde G_3)_{11} & = &
\frac{\tau^2}{3 z^2} \,
\Big [z (4 s v - s^2 - v^2) - 6 s v \tau \Big ] ,
\nonumber \\
(\tilde G_3)_{12} & = &
(\tilde G_3)_{21}  = \frac{s \tau}{3 z^2} \,
\nonumber \\
& \times &
\Big [ v^3 + s^3 + \tau^3 + 3 s \tau v - z (s^2 + \tau^2) \Big ] ,
\nonumber \\
(\tilde G_3)_{22} & = &
\frac{s^2}{3 z^2} \,
\Big [z (4 v \tau - \tau^2 - v^2) - 6 s v \tau \Big ] .
\nonumber
\end{eqnarray}
Here the variable $z$ is defined as $z = s + v + \tau$.

\section{Expansion in the basis in an external
         electromagnetic field}
\label{app:basis}

Calculations are significantly simplified by using
a basis which is constructed of the photon four-momentum~$q_\mu$
and the external electromagnetic field tensor
$F_{\mu \nu}$~\cite{Ritus}:
\begin{eqnarray}
b_\mu^{(1)} & = & (q F)_\mu, 
\qquad 
b_\mu^{(2)} = (q \tilde F)_\mu, 
\label{eq:basis} \\
b_\mu^{(3)} & = & q^2 (q F F)_\mu - q_\mu (q F F q),
\qquad 
b_\mu^{(4)} = q_\mu.
\nonumber
\end{eqnarray}
In the specified basis a polarization vector of an arbitrary photon is 
\begin{eqnarray}
 \varepsilon_\mu & = & \frac{(q \varepsilon)}{q^2} \, q_\mu -
\frac{(Ff)}{2 (qFFq)} \,
 (Fq)_\mu - \frac{(\tilde F f)}{2 (qFFq)} \, (\tilde F q)_\mu
\label{eq:EP} \\
 & - & \frac{(qfFFq) }{ q^2 \, (qFFq)^2} \,
 \Big [ q^2 \, (FFq)_\mu - (qFFq) \, q_\mu \Big ],
\nonumber
\end{eqnarray}
where $f_{\alpha \beta} = q_\alpha \varepsilon_\beta -
 q_\beta \varepsilon_\alpha$ is the photon field tensor.
Note that the last term in Eq.~(\ref{eq:EP})
is not equal to zero only for virtual photons.

Taking into account Eq.~(\ref{eq:EP}), we can reduce all structures
of the type $(l_1 f l_2)$ that appear in the expression for the vertex
function $\Lambda$ to the form:
\begin{eqnarray}
   (l_1 f l_2) & = & (f F) \,
\frac{(l_1 q) (q F l_2) - (l_2 q) (q F l_1) }{2 (qFFq)}
\label{eq:LF1} \\
   & + & (f {\tilde F}) \, \frac{(l_1 q) (q {\tilde F} l_2) -
   (l_2 q) (q {\tilde F} l_1) }{2 (qFFq)}
\nonumber \\
   & + & \frac{(l_2 q) (q f F F l_1) - (l_1 q) (q f F F l_2)}{(qFFq)},
\nonumber
\end{eqnarray}
where $l_1$ and $l_2$ are arbitrary four-vectors.
Below we give some structures for which we used the
representation~(\ref{eq:LF1}):
\begin{eqnarray}
(q_1 f_2 F F q_1) & = &
\frac{(q_1 F F q_2)}{2 (q_2 F F q_2)} \,
\label{eq:LF2} \\
& \times &
\Big [ (f_2 F)(q_1 F q_2) + (f_2 \tilde F)(q_1 \tilde F q_2) \Big ],
\nonumber \\
(q_1 f_2 F F q_2) & = & \frac{1}{2} \,
\Big [ (f_2 F)(q_1 F q_2) + (f_2 \tilde F)(q_1 \tilde F q_2) \Big ],
\nonumber \\
(q_1 f_2 F q_2) & = &
\frac{1}{2} \, (q_1 q_2) (f_2 F),
\nonumber
\end{eqnarray}
where $(f_i)_{\alpha \beta}$ is the electromagnetic field tensor
of the $i$th photon ($i = 1, 2$).

Note that the following relationship:
\begin{equation}
(T_1 \tilde T_2)_{\mu \nu} + (T_2 \tilde T_1)_{\mu \nu} =
\frac{1}{2} \, (T_1 \tilde T_2) \, g_{\mu \nu}
\label{eq:T1T2relation}
\end{equation}
proves very helpful between two arbitrary antisymmetric
tensors $(T_1)_{\mu \nu}$ and $(T_2)_{\mu \nu}$. If for tensors
$T_1$ and $T_2$ we take the constant electromagnetic
field tensor $T_1 = T_2 = F$, which meets the condition
$(F \tilde F) = 0$, then from Eq.~(\ref{eq:T1T2relation})
the condition on the electromagnetic field tensor $F$ follows
automatically,
\begin{displaymath}
(F \tilde F)_{\mu \nu} = 0.
\end{displaymath}

\end{document}